# Implementing structural slow light on short length scales: the photonic speed-bump

REMI FAGGIANI[1], JIANJI YANG[1,#], RICHARD HOSTEIN[2] AND PHILIPPE LALANNE[1,*]

[1]*Laboratoire Photonique, Numérique et Nanosciences (LP2N), UMR 5298, CNRS-IOGS-Univ. Bordeaux, Institut d'Optique d'Aquitaine, 33400 Talence, France.*
[2]*Sorbonne Universités, UPMC Univ Paris 06, CNRS UMR 7588, Institut des NanoSciences de Paris, F-75005, Paris, France.*
[#]*Now at Department of Electrical Engineering, Stanford University, Stanford, California 94305, United States*
[*]*Corresponding author: philippe.lalanne@institutoptique.fr*

**One-dimensional (1D) infinite periodic systems exhibit vanishing group velocity and diverging density of states (DOS) near band edges. However, in practice, systems have finite sizes and inevitably this prompts the question of whether helpful physical quantities related to infinite systems, such as the group velocity that is deduced from the band structure, remain relevant in finite systems. For instance, one may wonder how the DOS divergence can be approached with finite systems. Intuitively, one may expect that the implementation of larger and larger DOS, or equivalently smaller and smaller group velocities, would critically increase the system length. Based on general 1D-wave-physics arguments, we demonstrate that the large slow-light DOS enhancement of periodic systems can be observed with very short systems, whose lengths scale with the logarithm of the inverse of the group velocities. The understanding obtained for 1D systems leads us to propose a novel sort of microstructure to enhance light-matter interaction, a sort of photonic speed bump that abruptly changes the speed of light by a few orders of magnitude without any reflection. We show that the DOS enhancements of speed bumps result from a classical electromagnetic resonance characterized by a single resonance mode and also that the nature and the properties of the resonance are markedly different from those of classical defect-mode photonic-crystal cavities.**

## 1. Introduction

Introducing a 1D periodic modulation into an initially uniform wire results in a bandgap opening and a redistribution of the states that cluster at the band edge, leading to a divergence of the density of states (DOS), known in solid-state physics as the Van Hove singularity [1]. The DOS singularity and the related group-velocity reduction have important consequences in optics. They are responsible for the feedback mechanism in distributed-feedback solid-state lasers [2,3], help implementing nonlinear processes [4-6], enhance the mode lifetimes of photonic-crystal cavities [7,8] and the sensitivity of optical sensors [9,10], and serve as key building blocks for interfacing light with atoms for a range of applications in photonic quantum-information processing [11,12].

The singularity (Fig. 1a) exists only in infinite systems, meaning that an infinite number of unitary cells are required to build it up. Intuitively, if one assumes that a resonance state can be attached to every individual cell or a few cells, the singularity is seen as originating from a coherent superposition of an infinite number of resonance states, which are all phase-matched and form a Bloch state with a null group-velocity at a precise frequency. However, no structure is strictly periodic and in any real device, e.g. photonic-crystal cavities [13], single photon sources [14], lasers [15], slow waves are reflected at the device termination and since the reflectivity considerably increases as the group velocity vanishes [16], only very weak reminiscences of the DOS singularity are observed.

Slow-light DOS enhancements are always entangled with other cavity-like DOS effects [7,8,16-19], so that the Van Hove singularity of on slow-light-assisted spontaneous-emission rate in periodic waveguides and cavities [14], it is not clear if the DOS enhancement associated to given slowness can be observed in a finite-length structure, especially for large slowness. Nor it is evident what is the minimum number of cells required to experience the slowness, or how close can one approach the DOS singularity with a finite structure. The goal of this work is to provide the answers to these questions and to propose a new structure to observe slow-light DOS enhancements.

We first consider 1D periodic systems. Based on general wave-physics arguments, we explain how to engineer the impedance at the system terminations to achieve slow-light DOS enhancements on very short lengths that scale with the logarithm of the inverse of the group velocities, the group index. Then in a second step, capitalizing on the mature knowledge recently gained on photonic-crystal (PhC) waveguides, we validate the previous general considerations with realistic designs. This leads us to propose a totally new family of photonic micro-resonators, photonic analogues of speed-bumps, in which the electromagnetic energy accumulates, not because of a resonant recirculation between two mirrors, but because of a sudden reduction of the group velocity, followed by a reciprocal acceleration to go back to the initial speed. 3D simulations of the transmission and DOS enhancement conclusively support that short systems may mimic the slow-light DOS enhancements of infinite systems even for very small group velocities.

## 2. 1D toy-model

Let us start by considering a 1D periodic system (Fig. 1a) composed of alternating layers of high and low indices for instance. In 1D, only a the gap) or propagative (in the band), exists, and the DOS exhibits the classical $(\omega - \omega_0)^{-1/2}$ divergence at the band edge $\omega_0$. For a periodic system of finite length $L$, Bloch modes are back-reflected at the terminations, the singularity is smoothed and the DOS in the bandgap is no longer zero [20]. Intuitively, if one assumes that the momentum space is typically sampled with a resolution of $1/L$ because of the finite length, one expects to observe group velocities $v_g$ that scales inversely proportional to the system length, $v_g \propto 1/L$, see Fig. 1b.

Actually, much shorter length scales can be achieved with suitably engineered terminations. The local DOS can be evaluated by considering the total power radiated by a Dirac dipole source at a working frequency $\omega$. The radiation mechanism in 1D is very simple. The source, a 2D current sheet, excites the two counter-propagative Bloch modes existing at $\omega$. The modes propagate until they reach the terminations. Thus if the terminations are equipped with tapers whose impedance matches the slow Bloch modes with the fast modes of the surrounding medium (Fig. 1c), no back-reflection occurs at the terminations. The tapers act as trompe-l'oeil that mimics the infinite system: the source sees the same impedance in the finite or infinite systems and radiates identically.

Counter-intuitively, our simple reasoning also suggests that the number of periods of the finite-length crystals is unimportant, and might be 1,2 or many, since there is no back-action of the terminations on the source. Thus answering our initial question of how many cells are required to experience a given slowness $v_g$ amounts to answer what is the minimum length required to *perfectly* bridge the impedance mismatch between a slow Bloch mode and a fast mode. Adiabatic tapers based on a gradual variation of the geometry could be a first alternative [21]. Such tapers have the advantage to offer broadband operation and to conform to any kind of waveguide geometries, but their lengths may become prohibitively long at small $v_g$'s. For 1D systems, it is possible to design much shorter tapers. By assuming that slow-Bloch modes of 1D periodic media are stationary patterns locally formed by two counter-propagating plane waves, *perfectly-matched* tapers formed by a combination of a phase plate and a quarter-wave stack can be designed at any arbitrarily-small $v_g$ [22]. Such tapers are not broadband, but interestingly offer small lengths that scale with the logarithm of the group index, $L \sim log(n_g)$ with $n_g = c/v_g$. This implies that perfect matching at $n_g = 10^4$ for instance is achieved with a taper length as small as four quarter-wave periods.

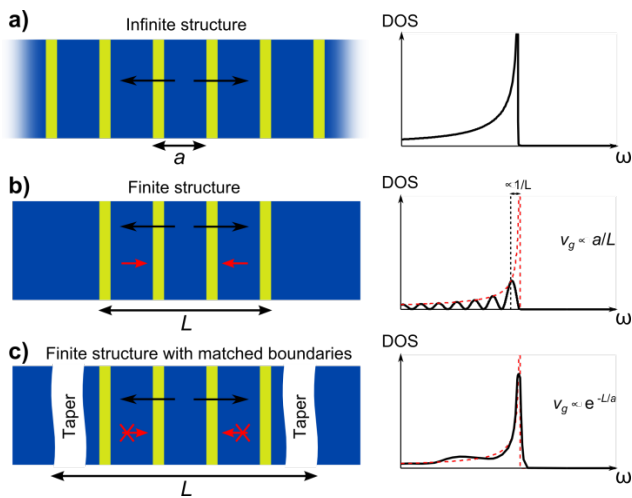

**Figure 1. Mimicking periodic infinite media with finite periodic media. a)** 1D infinite periodic medium: the source couples to outgoing Bloch modes (black) that propagate away without any reflection. **b)** 1D finite system: the Bloch modes are back-reflected on the terminations and modify the field distribution inside the periodic structure. **c)** 1D

sees an infinite periodic medium. **a)-c)** The right insets sketch the corresponding DOS.

Albeit simplistic, the 1D toy model suggests that the DOS enhancement at the band edge of periodic media, which is generally attributed to slow Bloch modes in infinite media, may be recovered in a very short length scale, $L \propto log(n_g)$. In a different point of view, one may illuminate the tapered system from outside with a plane wave, rather than with an internal source. The structure then can be seen as the analogue of a "speed bump" for photons. The light propagating at large group velocity in the surrounding medium is first slowed down in the taper and then propagates at slower group velocity in the periodic system, before being reaccelerated in the second taper and escaping in the surrounding medium on the opposite side. Indeed, the perfectly-matched impedance ensures a complete transmission at the working frequency $\omega$.

Structural slow light has many applications in science and technology, from delay lines [23], pulse compression [24] and sensors [9,10] to single-photon components for quantum information [14]. Hereafter, we implement the concept of photonic speed bumps, developed in 1D, with the aim to design a manufacturable photonic structure that potentially offers new properties.

### 3. Photonic-crystal speed bump

We will rely on the mature photonic crystal (PhC) waveguide platform to study the optical properties of realistic speed bumps. A schematic of the proposed speed bump is shown in Fig. 2a. A finite-length *N*-period-long slow-W1 waveguide (with one row of holes missing in the $\Gamma M$ direction of the photonic lattice) is bridged, by two tapers, to two semi-infinite fast-W1 waveguides obtained by slightly elongating the period of the slow-W1 waveguide in the longitudinal direction. This structure will be referred to as the *N*-period long speed bump. The geometrical parameters of the waveguides are given in the caption of Fig. 2.

Hereafter, the theoretical analysis is performed with a 3D fully-vectorial Fourier-modal method [25], which relies on an analytical integration of Maxwell's equations along the waveguide direction through a Bloch mode expansion of the field. The method has already been successfully applied to accurately analyze various scattering problems in photonic crystal waveguides and in sequences of them, see [26,27] for instance.

The dispersion curves of the slow- and fast-W1 Bloch modes are plotted in Fig. 2b. In the spectral range of interest, close to the band edge of the slow-W1 waveguide marked by the horizontal black dotted line, the waveguides are monomode. The slow-W1 waveguide exhibits a small group velocity, while the fast-W1 waveguides exhibit rather large and nearly constant group velocity $c/6$. Figure 2c shows the normalized amplitude of the *y*-component of the magnetic field in the median plane of the membrane at $\omega/c = 6.816 \mu m^{-1}$, $n_g = 6$ and 1000 for the fast- and slow-W1 waveguides respectively. The mode profiles and amplitudes are sharply distinct due to the group-velocity impedance mismatch. We note that the fast-W1 waveguides may be replaced with other types of waveguides, e.g. ridge waveguides; the choice just impacts the taper design, but not the principle of operation of the speed bump.

The impedance mismatch leads to large modal reflectance at the interface between the slow- and fast-W1 waveguides, see the black curve in Fig. 3b. Thus the PhC speed bump design solely and critically relies on our ability to suppress the back reflection by effectively tapering the fast- and slow-W1 waveguides to mimic an infinite slow-light system. To design the tapers, several geometries and optimization techniques can be considered [28,29]. Hereafter, we follow the approach in [26].

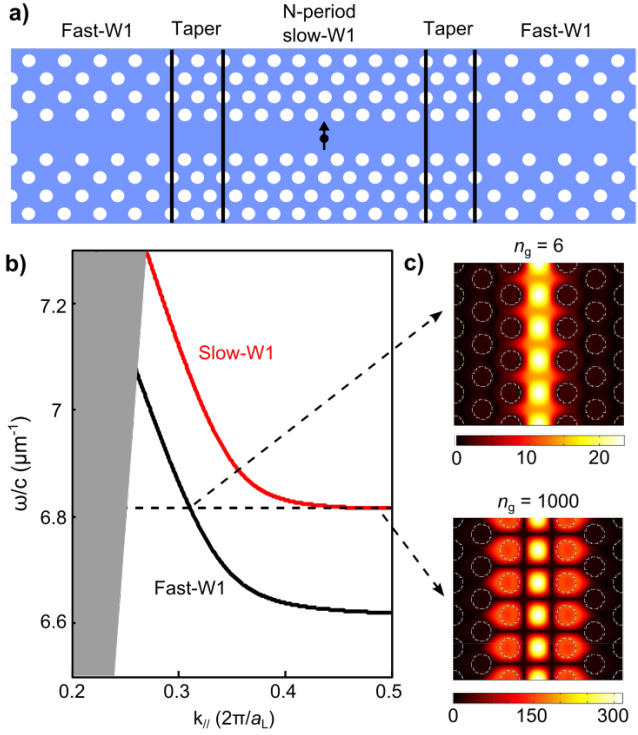

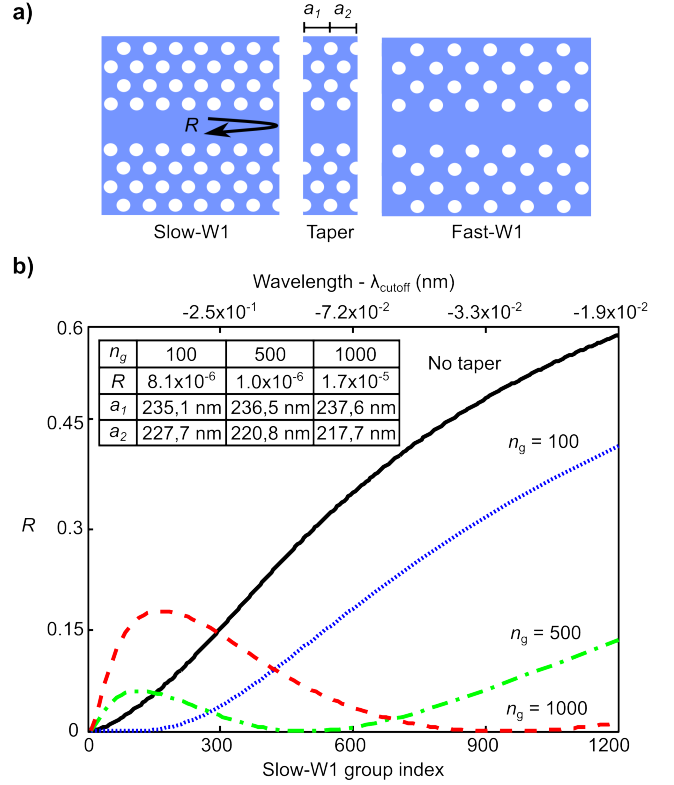

**Figure 2. Photonic-crystal speed bump. a)** Schematic of an $N$-period-long speed bump, composed of an $N$-period-long slow-W1 waveguide surrounded by bilayer heterostructure tapers and fast-W1 waveguides. The waveguides are assumed to be etched in a membrane of thickness $220nm$ and refractive index 3.45. The 2D photonic crystal mirrors of the slow-W1 waveguide are made of a triangular lattice of air holes of periodicity $a = 232nm$. The fast-W1 waveguide is obtained by stretching the longitudinal period of the slow-W1 waveguide from $a = 232nm$ to $a_L = 245nm$, the transverse period being unchanged. **b)** Dispersion curves of slow-W1 (red) and fast-W1 (black) waveguides. The operating frequency is marked with a black horizontal dashed line. **c)** Amplitude of the y-component of the magnetic field of the slow-W1 ($n_g = 1000$) and fast-W1 ($n_g = 6$) Bloch modes at the operating frequency. The modes are normalized to carry a power flow of 1.

Figure 3a shows the taper geometry. The latter is composed of a bilayer-heterostructure formed from the slow-W1 waveguide by slightly varying the longitudinal periods, $a_1$ and $a_2$. Similar bilayer tapers have been previously used to design remarkably short and effective couplers with 2D computational results [26], and have been used in many slow-light experiments afterwards [4,14,30]. The following numerical results are obtained with a 3D fully-vectorial method that includes scattering into the air clads. They evidence that the bilayer geometry is versatile since it allows us to design effective tapers from $n_g = 4$ up to 1000. Figure 3b shows the performance of 3 tapers optimized for 3 target group indices, $n_g = 100$ (dotted-dashed green curve), 500 (dotted-blue curve) and 1000 (dashed-red curve), respectively. In contrast with the abrupt interface case (no taper), the reflectance $R$ shows a markedly different behavior with a nearly-zero reflectance around the targeted $n_g$. Additionally, the calculations show that the scattering in the cladding is negligible and that the transmission is equal to 1-R with a very good approximation ($1 - R - T = 0.012$ for $n_g = 1000$). Although compactness comes at a price of narrowband operation, the ultra-short bilayer-heterostructure taper provides a very effective approach to reduce the impedance mismatch and to mimic an infinite system on short length scales. Also, it is worth noting that, the taper design presented here is just a proof-of-concept demonstration. Tapers with more than two achieve even better performances, in terms of, e.g., bandwidth, target group index, and tolerance to fabrication imperfection.

**Figure 3. Taper optimization. a)** Taper layout. A slow-W1 waveguide is connected to an elongated fast-W1 waveguide through a bilayer-heterostructure taper, which is optimized by tuning the longitudinal periods $a_1$ and $a_2$ of the two layers. **b)** Spectral dependence of the reflectance of the tapered interface for 3 tapers optimized to achieved an ultra-small reflectance (see the Table inset) for $n_g = 100$ (dotted blue), 500 (dotted-dashed green) and 1000 (dashed red) as a function of the group index of the slow W1 waveguide (bottom horizontal axis) or wavelength (top horizontal axis). The solid black curve shows the reflectance without taper. Inset: Minimum taper reflectance and taper geometrical parameters.

## 4. Speed bump LDOS

The capability of PhC speed bumps to mimic the DOS singularity is analyzed by computing the local DOS, or LDOS, seen by a $x$-polarized electric dipole (small arrow in Fig. 2a) placed in the center of the PhC speed bump. This quantity, normalized by the emission of the same dipole in the bulk, is commonly known as the Purcell factor. Figure 4 shows the normalized LDOS for a long speed bump ($N = 8$) for the optimized tapers of Fig. 3. For comparison, we also show the LDOS of the same emitter placed in an infinite slow-W1 waveguide (black dotted curve showing the van Hove singularity) and in a speed bump without taper (black solid curve).

The results evidence that the LDOS is substantially enhanced by tapering and that the enhancement is stronger for speed bumps with tapers that are designed for perfect impedance matching at high group indices. We also note that the LDOS is not null in the gap of the slow-W1 waveguide, simply because of the tunneling effect associated to the evanescent gap Bloch modes that are reflected at the taper interface. More interesting are the LDOS enhancements achieved at every target group indices, which are highlighted by circle markers. It is noteworthy that all markers are superimposed with the black-dashed curve obtained for the truly infinite slow-W1 waveguide, implying that, at their target group indices, the tapers act as real trompe-l'oeil and the mimic, at least in a narrow spectral band, the DOS of infinite periodic systems up to $n_g \approx 1000$ with compact systems of a few periods long. Let us specify that the frequency corresponding to the target group index, for which the taper is optimized, does not necessarily correspond to the maximum of the speed bump LDOS as shown in all curves.

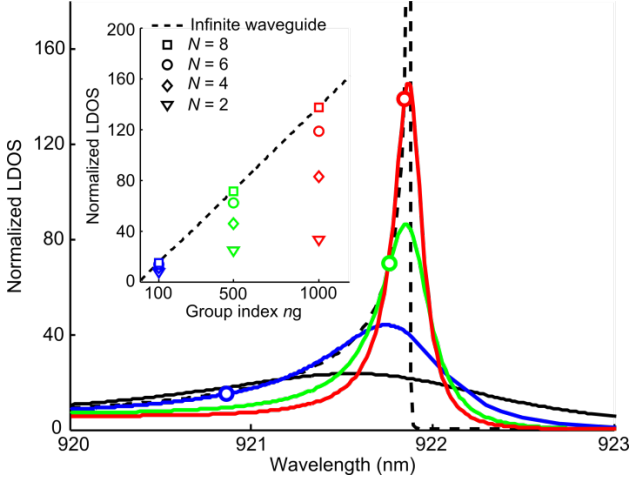

**Figure 4. Mimicking the van Hove singularity with PhC speed bumps.** Normalized LDOS (or Purcell factor) seen by an $x$-polarized source placed in the center of a 8-period-long speed bump optimized for operation at $n_g = 100$ (blue), 500 (green), 1000 (red) and without taper (black). The black dashed curve corresponds to the LDOS achieved for a fully-periodic, infinite slow-W1 waveguide (Van Hove singularity). The blue, green and red circles highlight the Purcell values achieved by the speed bump at their nominal operation wavelengths for which the taper reflectance is almost null. Importantly, the dots are almost superimposed with the black dashed curve, evidencing that the source emitting in the speed bump emits as if it were in a fully periodic waveguide. Inset: Evolution of the Purcell factor with speed bump length $Na$ for the same three optimized tapers at $n_g = 100, 500$ and $1000$.

The inset of Fig. 4 displays the LDOS enhancements achieved at the target group indices for several values of the speed bump length $N$. In contrast with the predictions of the 1D toy-model, the normalized LDOS vary with $N$. Further computations, not reported here, have shown that the LDOS at the target group index remains stable for $N > 8$. Unlike 1D thin-film stacks, PhC waveguides support, in addition to the guided Bloch modes, a few evanescent Bloch modes [26]. The source emission feeds all the modes. The latter propagate outward until reaching the tapers, where they scatter back (only the truly guided Bloch mode is fully transmitted) and potentially excite the inward-propagating guided Bloch modes, resulting in a back-action on the source. Overall, these intricate scattering processes involving evanescent Bloch modes not present in truly 1D systems may lead to a substantial change of the LDOS, preventing the observation of a mere signature of the slow-light LDOS enhancement. However, when the slow-W1 section length increases, the impact of evanescent modes fades away as they effectively lose energy before reaching the taper. More details are provided in the Supplemental Document.

## 5. Nature of the speed bump resonance mode

In classical defect-modes PhC microcavities [13], the resonance modes are essentially Fabry-Perot resonances formed by the bouncing back and forth of light in a region surrounded by two high-$R$ mirrors. In speed bumps, the trapping mechanism is different. The electromagnetic energy accumulates because of a sudden and drastic reduction of the light group velocity from $c/6$ to $c/1000$, followed by accumulation does not rely on a confinement with high-$R$ mirrors, but rather on the possibility for the light to efficiently escape the slow-light region. The speed bump can thus be seen as the photonic analogue of plasmonic nanofocusing devices, which provide strong field enhancements with slow plasmons in tiny air gaps formed at the mouth of almost touching metallic dimers [32,33].

Due to the unusual nature of the trapping mechanism, the question arises as to whether the DOS enhancement in speed bumps results from the excitation of an electromagnetic resonance. Often asymmetric lineshapes that are seemingly Lorentzian, such as those in Fig. 4, are not due to a single resonance but to an interference between two modes with very similar energies [34], see the striking example of Fig. 3 in [35]. To clarify the origin of the asymmetric lineshapes, we compute the resonance modes supported by the speed bump designed for a target group index of 100 in the spectral range of interest, around the band edge frequency $\omega_0 = 2.043 \cdot 10^{15} s^{-1}$. We find a single resonance mode (see Fig. 5a) with a complex eigenfrequency $\omega = \omega_0(0.999 - 0.001i)$ corresponding to a quality factor $Q = 500$. The mode volume, normalized with the method in [36], is $V = 0.092 + i0.039 \, \mu m^3$.

The quality factor $Q$ is related to the time spent by the electromagnetic wave in the structure. As shown by numerical results not reported here for the sake of clarity, it varies linearly with the speed bump length and the taper group index, as expected.

The mode volume is more interesting. Its real part, $\approx 0.1$ wavelength cube is comparable to those of classical PhC cavities. It directly quantifies the normalized LDOS enhancement (or Purcell factor $P(\omega)$) at resonance $\omega = \omega_0$, while the imaginary part, which accounts for the dispersion in the structure, is responsible for the asymmetry of the spectral lineshape response of $P(\omega)$ [35]. Figure 5b shows the normalized LDOS seen by a $x$-polarized emitting dipole placed in the center of the speed bump. The single-mode model prediction computed with the analytical expression for $P(\omega) = \frac{3}{4\pi}\left(\frac{\lambda_0}{n}\right)^3 \text{Re}\left(\frac{Q}{V}\right)\frac{\omega_0^2}{\omega^2}\frac{\omega_0^2}{\omega_0^2+4Q^2(\omega-\omega_0)^2}\left[1 + 2Q\frac{\omega-\omega_0}{\omega_0}\frac{\text{Re}(V)}{\text{Im}(V)}\right]$ is shown with the red-dashed curve. The model faithfully agrees with fully-vectorial Green-tensor calculations (black-solid curve), evidencing that the speed bump physics is governed by a single-electromagnetic highly-dispersive resonance.

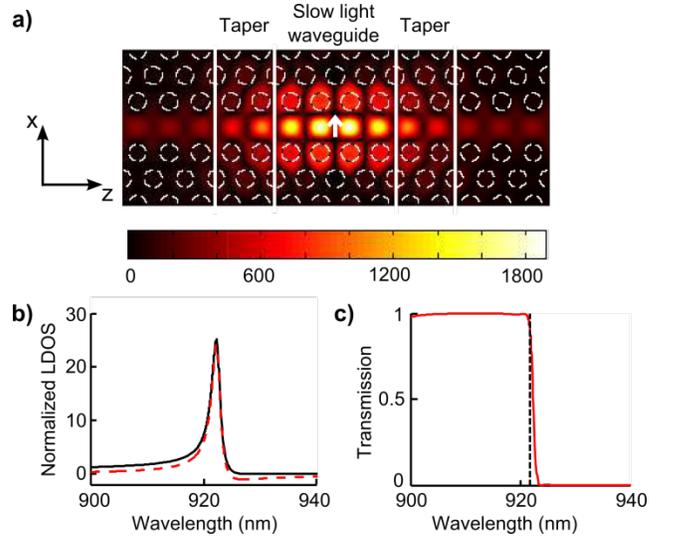

**Figure 5. Optical properties of the speed bump resonance mode. a)** $|H_y|$ for the resonance mode of a 4-period long speed bump. **b)** Normalized LDOS seen by the $x$-polarized electric dipole source placed in the center of the speed bump, see **a**). The black and dashed-red curves are respectively obtained with fully-vectorial Green-tensor calculations [27] and with the single-mode expansion formula. **c)** Step-like transmission (red curve) under illumination by the guided Bloch

band edge of the slow-W1 mode. All the results in a)-c) are obtained for a speed bump with a taper optimized for $n_g = 100$.

As shown by Fig. 5a, the resonance mode of speed bumps much resembles those of PhC microcavities.. This resemblance raises again the recurrent issue of the analogy and difference between cavity resonances and slowly traveling waves. The analogy stems from the fact that slow Bloch modes, just like cavities, result from two counter-propagative progressive waves that form an almost stationary pattern. Thus the speed bump may be seen as a stationary cavity, considering that the unidirectional flow of energy in the inner section results from two counter-propagating travelling waves with slightly different amplitudes. Following the analogy which amounts to consider a change of basis, one may intuitively understand that the taper that does not reflect light in the slow-Bloch-mode basis, acts as an efficient mirror in the progressive-wave basis, reflecting the incident waves into a counter-propagating one with a slightly smaller amplitude. More details on the analogy between 1D slow-light tapers and mirrors can be found in [22].

This intuitive reasoning well explains why the field distribution shown in Fig. 5a looks similar to the resonance modes of PhC cavities and why speed bumps can be well modeled by a single electromagnetic resonance. However, the boundary is subtle between resonances and slowly traveling waves, and as we will see now, speed bumps are not merely equivalent to PhC cavities, as they offer distinct properties and characteristics. This is essentially because slow Bloch modes are not a superposition of two independent travelling waves, but rather of two coupled waves with a distributed feedback.

A first substantial difference between speed bumps and cavities lies in the $\text{Im}(V)$ values. Comparatively, classical defect-modes PhC microcavities have much smaller $\text{Im}(V)$'s, and thus more symmetric lineshapes, than speed bumps. So far, large $\text{Im}(V)$'s have been observed only for plasmonic nanoantennas [35], for which the asymmetry of the DOS on the red and blue sides of the resonance is due to metal dispersion. For speed bumps, the asymmetry is not due to material dispersion, but to the highly dispersive nature of the slow-W1 Bloch mode at the band edge.

Another significant difference between speed bumps and cavities lies in the impact of the device length on the spectral properties. The spectral response of PhC microcavities strongly depends on the defect length, whereas the length of the slow-light region of speed bumps is only of minor importance. No precise phase accumulation and matching are required with speed bumps, as the resonance frequencies is in fact set by the cutoff frequency of the slow-W1 Bloch mode.

Yet another striking difference is shown in Fig. 5c. Unlike PhC microcavities that present an Airy-type symmetric Fabry-Perot transmission lineshape, the speed-bump transmission exhibits an unusual step-like lineshape, varying from 1 to 0 at the band edge of the slow-W1 waveguide with a steepness that slightly depends on the length of the slow-light region and the target group index used to design the tapers, see details in the Supplemental Document. A decisive advantage of the step-like lineshape over the Lorenzian resonance curve could be for implementing optical switches with high modulation contrasts.

## 6. Conclusion

We have added a new family of resonators to the long list of photonic microcavities. The photonic speed bumps are composed of a short slow-light section with a few identical unit cells that are impedance-matched with the outside space by short tapers. The latter give the impression to the slow wave that the short light section is much more extended and thus artificially enhance the DOS. The nature of speed bump resonances is markedly different from that of the resonances of classical defect-mode PhC microcavities. Speed bump resonances more resemble those found in plasmonic nanofocusing mirrors, but rather by implementing a sudden reduction of the light speed, followed by a reciprocal acceleration to go back to the initial speed.

Using the mature photonic-crystal-waveguide platform, we have designed realistic speed bumps that can be easily fabricated with present technologies. 3D simulations confirm the possibility of mimicking infinite periodic waveguides with compact devices that are only 12-period long and achieving large DOS enhancements corresponding to a group index of 1000. Just as for PhC microcavities, we have shown that the LDOS enhancement of speed bumps is due to a single electromagnetic resonance, but striking differences have also been observed between the two resonators. For instance, the resonance frequencies of speed bumps are no longer set by a Fabry-Perot-like phase-matching condition and are actually independent of the length of the slow-light periodic section. Also the transmission presents a highly unusual step-like character not encountered in PhC microcavities.

Photonic-crystal speed bumps may found applications for engineering photon emission and scattering. Owing to their capability to suddenly vary the light speed, these microscale dielectric devices are ideal candidates to implement 1D photon/emitter interfaces [37,38] capable of mediating long-range interactions between quantum emitters using photons propagating in fast guided modes and achieving nearly-perfect couplings between the emitters and the waveguide in slow light sections. In addition, the possibility of mimicking infinite periodic waveguides with compact devices may allow fundamental studies on structural slow light at ultra-low speeds, whose observation is usually hampered by localization effects due to inevitable fabrication errors [3,18]. In this regard, we provide a study of the impact of fabrication imperfections in the Supplemental Document and suggest that the DOS enhancements predicted for ideal speed bumps must be observable in real devices with lengths comparable to the localization length in W1 waveguides.

**Acknowledgment**. This work is partly supported by the French RENATECH network. RF acknowledges financial support from the French "Direction Générale de l'Armement" (DGA).

See Supplemental Document for supporting content

# Implementing structural slow light on short length scales: the photonic speed-bump

Rémi Faggiani,[1] Jianji Yang,[1, #] Richard Hostein,[2] and Philippe Lalanne[1,*]

[1]Laboratoire Photonique, Numérique et Nanosciences (LP2N), UMR 5298, CNRS-IOGS-Univ. Bordeaux, Institut d'Optique d'Aquitaine, 33400 Talence, France.
[2]Sorbonne Universités, UPMC Univ Paris 06, CNRS UMR 7588, Institut des NanoSciences de Paris, F-75005, Paris, France.
[#]Now at Department of Electrical Engineering, Stanford University, Stanford, California 94305, United States
*Corresponding author: philippe.lalanne@institutoptique.fr

**Content of the document**

(1) Impact of the evanescent Bloch modes on the speed bump LDOS
(2) Impact of disorder on the photonic speed bumps
(3) Transmission properties of the speed bumps

All the numerical data presented in the Supplemental Document are obtained with the 3D aperiodic Fourier Modal method, see main text for details.

## 1. Impact of the evanescent Bloch modes on the speed bump LDOS

W1 PhCWs support numerous evanescent modes in addition to their fundamental guided modes. Usually, these evanescent modes are neglected, owing to their small propagation lengths. However, they can have profound impact on the light transport on small length scale [Lec05,Mat09]. In the present speed bump, the resonant effect is dominantly due to the slowdown of the fundamental guided Bloch mode, indeed. However, evanescent Bloch modes also come into play.

They are directly excited by dipole sources in the LDOS computations, and then decay as they "propagate" towards the taper where they notably reflect back into the slow backpropagating Bloch mode, further perturbing the dipole source emission and enabling a bidirectional flow of energy in the slow PhCW section. They are also indirectly excited when the slow Bloch mode reaches the taper and then "propagate" back towards the source. Thus, if the length of the slow-W1 waveguide is not large compared to the decay length of the evanescent Bloch modes (for small $N$'s), a tunneling effect takes place in short speed bumps.

This is exactly what we observe here. The energy transfer in speed bumps for $N < 8$ periods is not solely due to the slow-W1 mode, and is perturbed by the evanescent modes.

The impact of evanescent Bloch modes is observed in the LDOS of speed bumps, see Fig. S1. As $N$ increases, the energy transfer through evanescent modes exponentially vanishes and the LDOS of the finite system tends towards that of the infinite system, for the group velocities (or energies) for which the tapers have been optimized. These energies are marked by small circles in the figure. This behavior is observed for the three tapers. We note that, for all cases, evanescent modes lower the LDOS, especially important for small $N$'s.

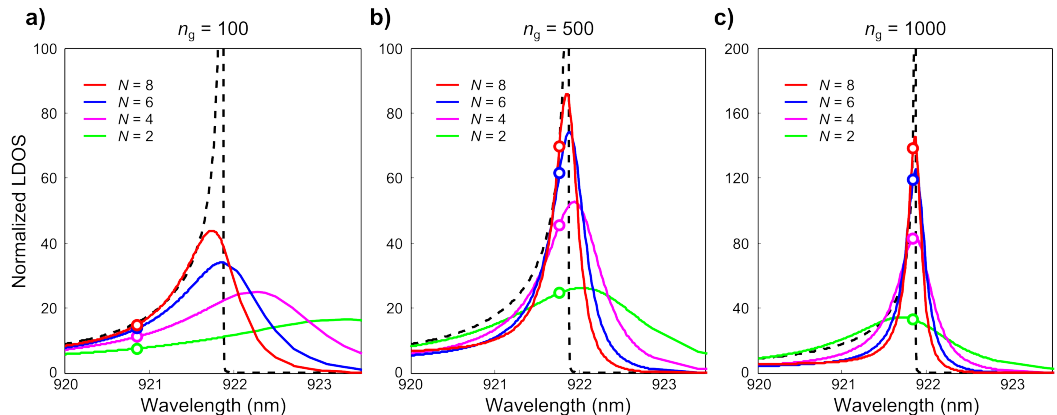

**Figure S1. Evolution of the Purcell factor with $N$ for the three tapers.** Normalized LDOS (or Purcell factor) seen by an $x$-polarized source placed in the center of speed bumps with $N$ = 2, 4, 6 and 8 periods and with tapers optimized for operation at $n_g = 100$ (**a**), $500$ (**b**) and $1000$ (**c**). The black dashed curves correspond to the LDOS achieved for a fully-periodic, infinite slow-W1 waveguide (Van Hove singularity). The green, pink, blue and red circles highlight the Purcell values for which the taper reflectance are almost null. Importantly, the red dots obtained for $N$ = 8 are almost superimposed with the black dashed curves, evidencing that sources in 8-period long speed bumps emit as if they were in a fully-periodic waveguide.

## 2. Impact of disorder on the speed bump performance

In this Section, we present a thorough theoretical study of the impact of fabrication imperfections on the DOS enhancement of photonic speed bumps using 3D fully-vectorial computations.

Fabrication imperfections are modeled by randomly varying the hole radii $\Delta R$ of the two inner-first rows of the W1 waveguide and of the tapers. In the fast W1 waveguides that surround the speed bump, the impact of disorder is negligible and the two inner row of holes are not randomly perturbed. This hypothesis further simplifies the numerical analysis. We assume that the $\Delta R$ values are independent from one hole to another and are null on average with a standard deviation $\langle \Delta R^2 \rangle = 1.5$ nm. This value is 2 to 3 times larger than state of the art technologies [Tag11], but has been proven to fairly predicts the performance of slow waveguides fabricated in laboratories equipped with second-class nanofabrication tools [Maz10].

We have computed the normalized local density of states (LDOS) spectrum at the center of the structure for 20 independent disorder realizations for *N* = 2,4 and 8 and for speed bumps optimized at $n_g = 100$ and 1000.

Figure S2 shows the results obtained for speed bumps equipped with tapers optimized for $n_g = 100$. For the sake of clarity, only 5 representative disorder realizations are plotted. The main conclusion is that the overall shape of the LDOS frequency-dependence is robust against disorder. The main deviation is a frequency-shift of the curves, especially stringent for *N* = 2. This inevitable (and expected) shift is simply due to the shift of the slow W1 mode cut-off frequency as the hole radii vary. For *N* = 8, the shift is less pronounced since, along the 8 periods, the slow mode experiences holes with either enlarged or reduced radii and then averaging occurs. However we note that the LDOS maximum enhancement present much stronger variations in comparison with the curves shown in (a) for small *N*'s. We additionally note that the normalized LDOS of perturbed speed bumps increases with *N*, just like for the ideal case.

On overall, the behavior of speed bumps operating with tapers optimized for $n_g = 100$ is found to be robust to fabrication imperfections, especially for small *N*'s.

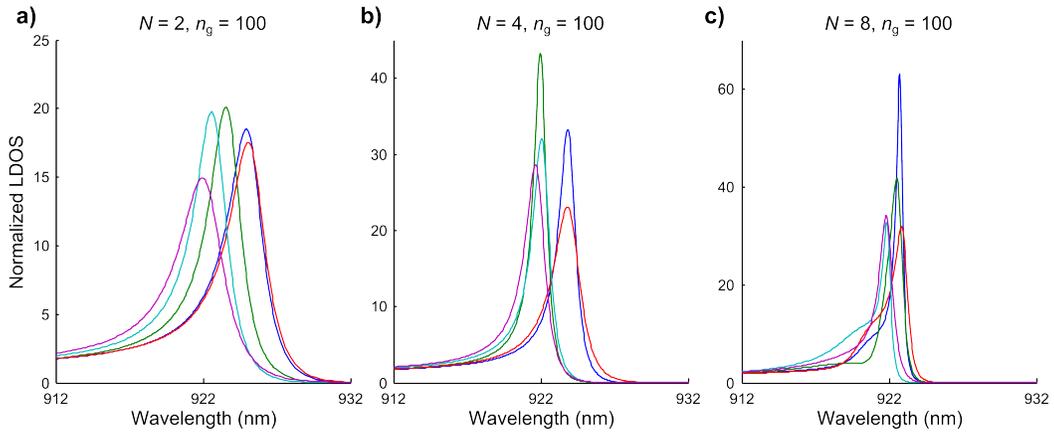

**Figure S2.** Normalized LDOS (or Purcell factor) seen by an *x*-polarized source placed in the center of a *N* = 2 (**a**), 4 (**b**) and 8 (**c**)-period-long disordered speed bump optimized at an operation frequency corresponding to $n_g = 100$. 5 typical LDOS response are shown for every *N* values.

As expected, the impact of fabrication imperfections is more stringent for speed bumps equipped with tapers optimized for $n_g = 1000$. The 2-period-long disordered speed bumps (Fig. S3a) is quite robust to disorder, except for the slight inevitable frequency shift already discussed. However, for increasing *N*'s, see Figs. 3b and especially 3c, huge variations of the maximum Purcell factor are observed. It is remarkable that for N = 8, the maximum values vary from ≈ 60 to nearly 1300. In our opinion, these large variations come from the resonant nature of the short tapers considered in this work, especially for large $n_g$ values. This may result from two effects: 1/ fabrication imperfections shift the optimized group index to smaller or larger values respectively resulting in a decrease or an increase of the LDOS, or 2/ the perturbed tapers becomes mirrors and classical cavity modes are replacing the speed bump. It would be interesting to study if the shape of the transmission curve is robust to disorder.

This leads us to the conclusion that the critical part of the present design is again the taper and that it is the taper that will limit the performance of the speed bump.

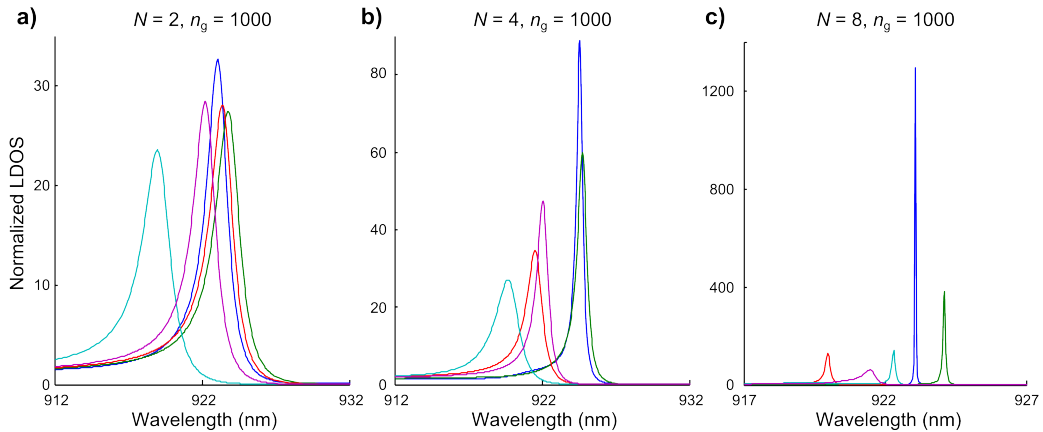

**Figure S3.** Normalized LDOS (or Purcell factor) seen by an *x*-polarized source placed in the center of a *N* = 2 (**a**), 4 (**b**) and 8 (**c**) -period-

## 3. Transmission properties

The transmission characteristic of speed bumps strongly differs from the Airy-function of classical microcavities, and displays a step-like transition at a frequency corresponding to the cutoff of the slow PhC waveguide.

For instance, the transmission $T$ of the 4-period-long speed bump optimized for operation at $n_g = 100$ (see Fig. 5c in the main text) undergoes a sharp transition from almost perfect transmission $T = 99.7\%$ to nearly zero $T = 0.02\%$ over a rather short spectral interval of 3 nm. This transition is due to the fact that the taper for $n_g = 100$ offers extremely small reflectivity for $n_g$ up to 150.

The behavior of the structure is however different for tapers optimized at larger $n_g$'s. Indeed, the resonant nature of the taper at large $n_g$'s leads to non-zero reflectivities for group-index values smaller than the values used for the design. This results into a narrow drop in the transmission for wavelength slightly smaller than the cutoff wavelength of the slow-W1 Bloch mode, as shown in Fig. S4.

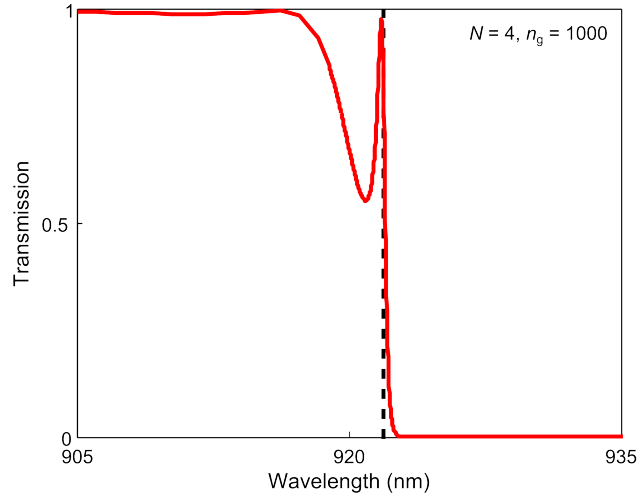

**Figure S4.** Transmission of a 4-period-long speed bump optimized for operation at $n_g = 1000$ under illumination by the guided Bloch mode of the fast-W1 waveguide. The black dashed line shows the band edge of the slow-W1 mode.